\documentclass[12pt,titlepage,preprint]{aastex}
\pdfoutput=1 

\usepackage[latin1]{inputenc}
\usepackage{amsfonts}
\usepackage{amsmath}
\usepackage{amssymb}
\usepackage{amsthm}
\usepackage[USenglish]{babel}
\usepackage{fontenc}
\usepackage{multicol}
\usepackage{upgreek}

\usepackage{pdfsync}
\usepackage{color}
\usepackage{latexsym}
\usepackage{epsfig}
\usepackage{xspace}

\usepackage[pdftex]{hyperref}

\newcommand{\ov}{\ensuremath{\mathbf{1}}\xspace}

\newcommand{\ev}{\ensuremath{\mathbf{e}}\xspace}

\newcommand{\xv}{\ensuremath{\mathbf{x}}\xspace}
\newcommand{\yv}{\ensuremath{\mathbf{y}}\xspace}

\newcommand{\xift}[3]{\ensuremath{x_{#1 #2}(#3)}\xspace}
\newcommand{\xif}[2]{\ensuremath{\mathbf{x}_{#1 #2}}\xspace}
\newcommand{\xit}[2]{\ensuremath{\mathbf{x}_{#1}(#2)}\xspace}
\newcommand{\xim}[1]{\ensuremath{\mathbf{X}_{#1}}\xspace}

\newcommand{\vectLength}[1]{\ensuremath{\left\| #1 \right\|}\xspace}

\newcommand{\cA}{{\ensuremath{\mathcal A}}\xspace}
\newcommand{\cC}{{\ensuremath{\mathcal C}}\xspace}

\newcommand{\cF}{{\ensuremath{\mathcal F}}\xspace}
\newcommand{\cH}{{\ensuremath{\mathcal H}}\xspace}
\newcommand{\cN}{{\ensuremath{\mathcal N}}\xspace}
\newcommand{\cR}{{\ensuremath{\mathcal R}}\xspace}

\newcommand{\cT}{{\ensuremath{\mathcal T}}\xspace}

\newcommand{\Ident}{\ensuremath{\mathbf{I}}\xspace}
\newcommand{\Onemat}{\ensuremath{\mathbf{J}}\xspace}

\newcommand{\Am}{\ensuremath{\mathbf{A}}\xspace}

\newcommand{\sC}{\ensuremath{\mathbb C}\xspace}
\newcommand{\sK}{\ensuremath{\mathbb K}\xspace}

\newcommand{\DM}{\text{DM}\xspace}

\newcommand{\notes}[1]{
    \textcolor{red}{\tt#1}}

\begin{document}

\title{Comparison of RFI Mitigation Strategies for Dispersed Pulse Detection}
\author{
John Hogden\altaffilmark{1} and
Scott {Vander Wiel}\altaffilmark{1} and
Geoffrey C. Bower\altaffilmark{2} and
Sarah Michalak\altaffilmark{1} and
Andrew Siemion\altaffilmark{2} and
Daniel Werthimer\altaffilmark{2}
}
\altaffiltext{1}{Los Alamos National Laboratory, Los Alamos, New Mexico, USA}
\altaffiltext{2}{University of California, Berkeley - Department of Astronomy, Berkeley, California, USA}

\begin{abstract}
Impulsive radio-frequency signals from astronomical sources are
dispersed by the frequency dependent index of refraction of the interstellar media and so appear as chirped signals
when they reach earth.  Searches for dispersed impulses have been
limited by false detections due to radio frequency interference (RFI)
and, in some cases, artifacts of the
instrumentation.  Many authors have discussed techniques to excise or
mitigate RFI in searches for fast transients, but comparisons between
different approaches are lacking.  This work develops RFI mitigation
techniques for use in searches for dispersed pulses, employing 
data recorded in a ``Fly's Eye'' mode of the Allen
Telescope Array as a test case.  We gauge the performance of 
several RFI mitigation techniques by adding
dispersed signals to data containing RFI and comparing false alarm
rates at the observed signal-to-noise ratios of the added signals.  
We find that Huber filtering is most effective at removing broadband
interferers, while frequency centering is most effective at
removing narrow frequency interferers.  Neither of these methods 
is effective over a broad range of interferers.  A method that
combines Huber filtering and adaptive interference cancellation 
provides the lowest number of false positives over the interferers
considered here.  The methods developed here have application to other 
searches for dispersed pulses in incoherent
spectra, especially those involving multiple beam systems.
\end{abstract}

\section{Background}
\label{sec:background}

Variable radio sources probe extreme physical conditions in the
Universe:  the sparks that emerge from high energy density regions
around compact objects such as black holes, neutron stars, and
magnetized stars, and the glowing embers that appear in the afterglow
of relativistic explosions \citep[e.g.,][]{cordes2004drs}. 
Fast transients are distinguished from slow transients both
physically and through the technology required to detect them.
Typically, fast transients originate from coherent emission
processes and have time scales of $\sim 1$ second or less.
Examples include pulsar emission, cyclotron masers, and
electrostatic discharges.  The short timescale of fast transients
drives a technological solution for discovery:  typically
these sources are found and characterized
through the analysis of high time resolution incoherent
spectra obtained from single dish telescope observations.

Pulsars are of great scientific interest.
These rotating neutron stars are the most accurate clocks in the Universe
and may be used for unique tests of general relativity, the nuclear
equation of state, and the processes of star formation and death 
\citep{2008ARA&A..46..541K}.  Pulsars produce both 
continuous pulse trains, that are discovered through periodicity searches,
as well as bright individual pulses, such as Crab giant pulses 
\citep{2003Natur.422..141H,2004ApJ...612..375C,2008ApJ...676.1200B}
and those from rotating radio transients \citep[RRATs][]{mclaughlin2006trb}.

Of significant interest is the recent discovery of a
very bright single pulse, only milliseconds in duration
\citep{lorimer2007bmr}. The pulse is inferred to originate outside of
the Galaxy, possibly at a distance of a billion light years, implying
a source of enormous energy density.  Subsequent investigations have supported both cosmological 
\citep{2011MNRAS.415.3065K}  and terrestrial origins 
\citep{2011ApJ...727...18B} for similar events.
The so-called ``Lorimer burst'' is controversial, however,
because of the possibility that the event is due to man-made radio signals
or radio frequency interference (RFI).

RFI presents a significant limitation
on the ability to detect and characterize pulsed emission.  
Man-made radio signals occur throughout the radio
spectrum, are variable in time and frequency, and can be strong enough
to be detected in the far-out sidelobes of the antenna primary beam
response \citep{ellingson2005introduction}. Examples of RFI are satellite
transmissions, aircraft communications, radar, TV, radio, cell phone,
and other point-to-point communication systems.  In the time domain, RFI may be
steady, erratic, repeating, or isolated and may have a broad range
of timescales.  In the frequency domain, RFI
may be narrowband, broadband, spread spectrum, regularly structured,
or irregularly structured.  Hybrids of these time- and frequency-modes,
such as swept-frequency signals, are common.

The dispersion of celestial pulses imposed by propagation 
through the ionized interstellar medium provides a unique signature
that is a powerful disciminant against RFI \citep[e.g.,][]{2009ApJ...703.2259D}.  
Nevertheless, detection methods can be improved through the use of algorithms that 
excise or mitigate RFI.  Standard searching for pulsed emission removes RFI
through excision of frequency channels and time segments that are
suspected to contain RFI based on 
amplitude thresholding \citep[e.g.,the PRESTO software,][]{2001AAS...19911903R}. 

In general, a wide range of methods for identification and mitigation of 
RFI has been considered \citep[most recently reviewed by][]{2010rfim.workE...1B},
including post-correlation matrix projection methods \citep{2000ApJS..131..355L,kocz2010radio},
blanking \citep[e.g.,][]{2009ApJ...703.2259D}
and coherent subtraction \citep{2003ApJS..147..167E}.
Recently, kurtosis in the distribution of voltage measurements
has been employed for RFI detection for single dish
data \citep[e.g.,][]{2010MNRAS.406L..60N}.
Each technique has strengths and weaknesses relative to different 
types of RFI.  Post-correlation methods are appropriate to interferometric
visibility data.  Blanking is most often applied to impulsive time-domain
RFI.
RFI rejection is carried out both in post-processing and in
real-time through dedicated 
digital instrumentation 
\citep[e.g.,][]{1997A&AS..126..161W}.

Of particular applicability to
pulse detection from single dish systems is  adaptive interference
cancellation \citep[AIC,][]{widrow1985adaptative} in which a small reference
antenna is employed for detection of a voltage stream that contains
the RFI but not the astronomical signal.  Cross-correlation of the two
voltage streams can  generate weights that are used to
subtract the reference stream from the astronomy stream.  The AIC
method was first used for RFI excision in radio astronomy by
\cite{1998AJ....116.2598B} and has been applied or is being considered
for many new radio telescopes \citep{kesteven2005afr,4566697}.
\cite{Bower2005Radio-frequency} discusses the theory of using AIC with
telescope arrays.  Laboratory and field tests have demonstrated that
the technique can effectively cancel interferers
\citep[e.g.,][]{bower2001application}.  AIC has an advantage over
other techniques, in that it makes no assumption about the frequency-
or time-domain characteristics of the RFI.

Searches for dispersed pulses should robustly cope with a wide
variety of interference, producing low false detection rates with
little impact on sensitivity to true astronomical impulses. 
Our goal is to develop and compare different RFI
mitigation techniques to help determine which are most useful.  In
particular, we are interested in exploring the efficacy of different
methods through an evaluation based on actual RFI observed in incoherent
spectra.  Theoretical estimates of algorithm performance can be
valuable, but it is almost always the case for RFI mitigation that 
the variety of RFI phenomena imply that no particular method is ideal 
or will meet its theoretical sensitivity under all (or any) circumstances.

In Section \ref{sec:flyseye}, we describe the
incoherent spectra data obtained in the ATA Fly's Eye experiment.
In Section \ref{sec:RFI},
we provide a compact mathematical formalism for the set of 
RFI filtering techniques that we are exploring in this paper.  We
do this with the goal of providing a clear statement of the content
of the techniques explored; many of these techniques have been
put to use by other researchers.  Since different methods of signal detection
may give different false alarm results when paired
with the same filtering method, in Section \ref{sec:detect} we
describe the chirp detection technique we used in our study of various
RFI filters.  Section \ref{sec:methods} describes how we tested
combinations of RFI filters with a chirp detection algorithm on various
forms of RFI.  Results of combining the mitigation techniques in
various ways are shown in Section \ref{sec:results}.  Conclusions are
in Section \ref{sec:summary-conclusions}.

Our results should guide further research in RFI mitigation for searches using incoherent spectra.  In particular,
the use of multi-beam systems such as those at Parkes and Arecibo can directly make use of 
the techniques described here.  Future instruments such as ASKAP may also make use of incoherent spectra
for detection of fast transients \citep{2010PASA...27..272M}, and these methods would be readily applicable there as well.

\section{Fly's Eye Data \label{sec:flyseye}}

We have carried out an observing campaign using the Allen Telescope
Array \citep[ATA,][]{welch2009allen} that uses a novel observing technique, described below,
to achieve high sensitivity to very bright, very rare, short-duration
transients.  
The Fly's Eye survey was carried out to detect events similar to the Lorimer burst.
The ATA consists of 42 6.1-m
dishes, each equipped with a log-periodic feed that is instantaneously
sensitive to radio frequencies from 0.5 to 11.2 GHz. 

Data from the ATA were captured in a fast spectrum 
fly's eye mode \citep{2010AcAau..67.1342S,Siemion2010The-Flys-Eye} in
which each antenna can be pointed to a different patch of the sky in order to cover a large area at the expense of interferometric information and thus spatial resolution.
In this mode, the digitally sampled waveform from each antenna in the array is mixed to baseband and
converted to the frequency domain (channelized) via Fourier techniques; in this case, a 128
channel streaming polyphase filterbank is computed over 512-sample windows.  Successive channelized windows are
accumulated as a power spectrum and the cumulative spectra are written
rapidly to disk.  We recorded 128-channel accumulated power spectra at a continuous 
rate of 1600 spectra per second for each of 44 antennas using 8 bits of precision. 
Data were obtained at RF frequencies that spanned from 1325 to 1535 MHz.
Accumulated power spectra are known as incoherent data because the
detection process removes phase information from the signal.

\begin{figure}[t]
  \centering
  \includegraphics*[width=2.4in]{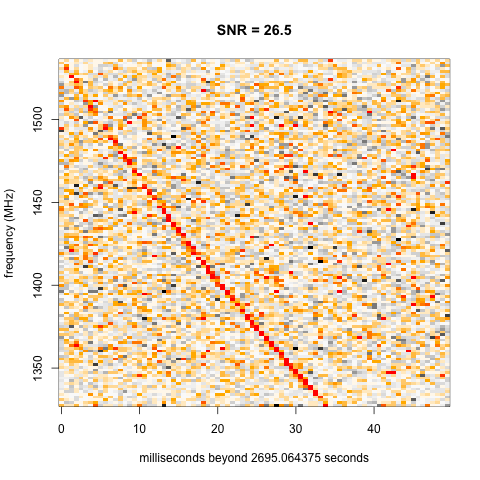}
  \includegraphics*[width=2.4in]{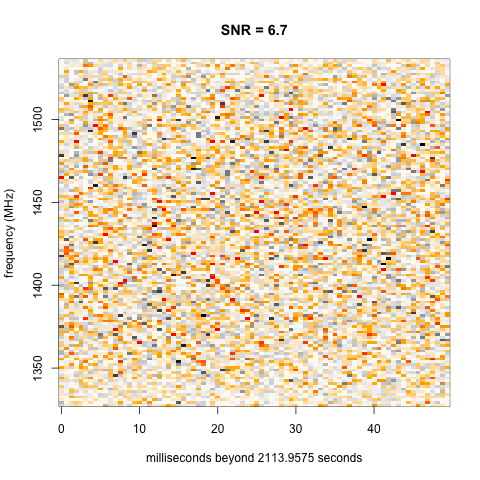}
  \caption{Two detections of giant pulses from the Crab pulsar.  The left
    plot shows the brightest pulse detected in one hour of observation
    using 44 input streams.  The right plot shows the 13th brightest
    pulse.  Intensity is represented by the pixel colors scale from red to black, with
red representing the highest intensity.}
  \label{fig:crabs}
\end{figure}

Figure~\ref{fig:crabs} shows two examples of dispersed pulses from
the Crab pulsar as observed in ATA fast spectrum data.  The left-hand
plot shows a strong detection, with signal to noise (SNR) of 26.5,
while the right-hand plot shows an event with SNR of 6.7.  The Crab pulsar has a dispersion measure of $\DM\sim57 \text{pc/cm}^3$, which translates to a 35 ms
delay between the highest and lowest frequency channels in the observed band. The curves
follow the expected quadratic dependence of the cold plasma dispersion
relation.

\begin{figure}[t]
  \centering
  \includegraphics*[width=.49\textwidth]{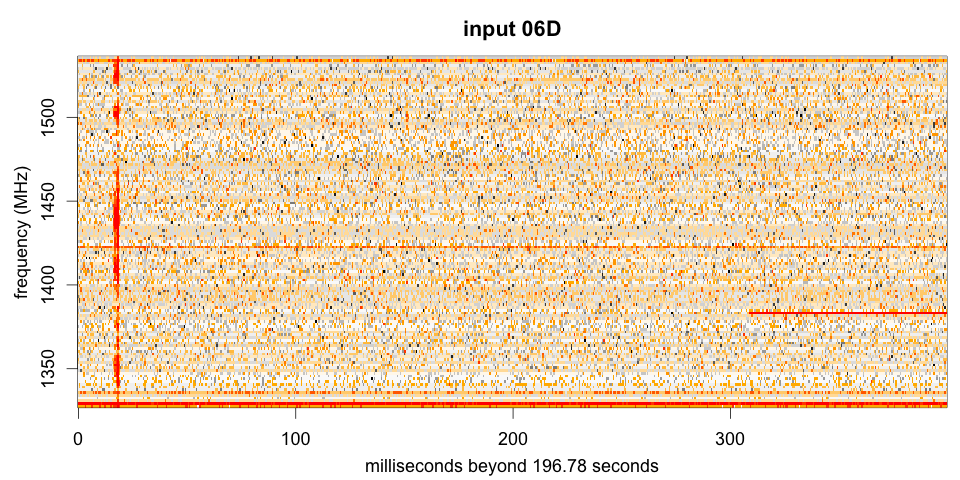}
  \includegraphics*[width=.49\textwidth]{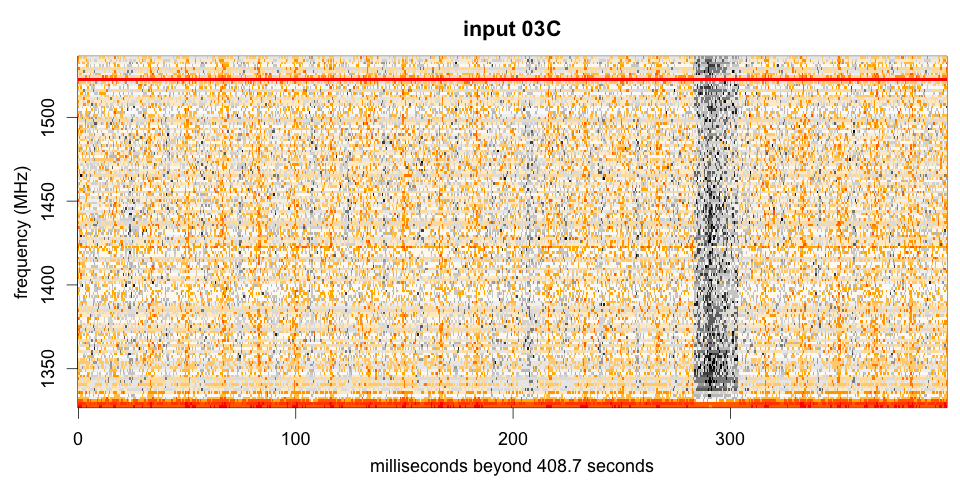}
 \caption{Examples of RFI in two spectrograms.  The left plot has an
   impulse across the full frequency range at 20 ms and a transient increase at 1380
   MHz beginning at 310 ms.  The right plot has a 60 cycle per second
   pattern across the frequency range and a 20 ms wide dark vertical feature
   ending at 300 ms.  Several horizontal stripes at various shades
   of red are frequency channels with more power than their
   neighbors.
   A stripe at 1425 MHz appears in both plots and indicates a
   persistently higher energy at that frequency.
   }
  \label{fig:mess}
\end{figure}

Figure~\ref{fig:mess} shows examples of RFI from the Fly's Eye data.  
Each image is a 
spectrogram of data collected on a single antenna at the ATA.  The
band pass response of the antenna (a smooth function of frequency) has
been subtracted from the spectra before plotting in order to highlight
the structure of the interference.  In the left-hand image, the
vertical stripe at 20 ms is an RFI impulse that affects all frequency
channels, whereas the horizontal segment beginning at 310 ms is a
transient RFI event at a single frequency, 1380 MHz.   The right-hand
image shows two prominent features; a 60 cycle per second pattern synchronized over all
frequencies and a 20 ms period of time, ending at 300 ms, in which all
frequencies have lower power than usual.  Several of the frequency
channels in either image have power levels that are sharply higher
than adjacent channels and subtle horizontal striations indicate that
power does not always vary smoothly with frequency. 
The goal of this work is to mitigate the effects of these and 
other types of RFI in anomaly detection in Fly's Eye data.

\section{RFI Mitigation Filters}
\label{sec:RFI}

Five RFI filtering methods are discussed below: {\em time centering} removes means for each time sample in a spectrogram whereas {\em frequency centering} removes means for each frequency; {\em energy clipping}
normalizes spectra whose total energy exceeds a threshold; {\em Huber
normalization} is a nonlinear high-pass normalizing filter in the time
direction that clips outlying pixels to $L$ standard deviations; and
{\em adaptive interference cancellation} (AIC) cleans the signal from a
target antenna by removing any correlated portion that it shares with
a set of reference antennas.  Each filter operates on a {\em
  spectrogram} (sequence of spectra).  In the case of AIC, a set of
reference spectrograms is also required.  Filters are sometimes used
in series, so each input spectrogram may have already been modified by
a previous filter.  The material introduced in this section draws on many sources; the goal here
is to present a compact formalism that permits us to compare the efficacy of
these methods.

The following notation is used in the remainder of this paper.
Let $\xift{i}{f}{t} \in \mathbb{R}$ denote the energy from input
spectrogram $i$ at frequency index $f$ for time sample $t$ with $i \in
\{1,\ldots,I\}$, $f \in \{1,\ldots,F\}$, and $t \in \{1,\ldots, T\}$. 

We use $\xif{i}{f}= \left[ \xift{i}{f}{1},  \xift{i}{f}{2}, \dots,
  \xift{i}{f}{T}  \right]$ (written as a row vector) to specify the
energy over time for frequency $f$ in spectrogram $i$.  The spectrum
at time sample $t$ from spectrogram $i$ is denoted by the column vector
\[\xit{i}{t}= \left[ 
  \begin{matrix}
      \xift{i}{1}{t} \\ \xift{i}{2}{t} \\  \vdots   \\ \xift{i}{F}{t}
  \end{matrix}
\right].\]

$\xim{i} = \left[ \xit{i}{1}, \xit{i}{2}, \ldots,  \xit{i}{T}  \right]$ is the $i$-th spectrogram.

\subsection{Time Centering, \cT}
\label{sec:time-centering}
Time centering refers to subtracting the mean of $\xit{i}{t}$ from
each of its elements.  This basic operation removes fluctuations in
total energy from one spectrum to the next.  After centering, the
(residual) energy in each spectrum sums to zero.  Time centering the
$i$-th spectrogram is defined as
\[ 
\cT \xim{i} \equiv \left(\Ident_F - \textstyle\frac{1}{F}\Onemat_{F}\right)\xim{i}
\]
where $\Ident_F$ is the identity matrix and $\Onemat_{F}$ is a square
matrix of ones, both matrices having dimension $F\times F$.  The
operator notation, $\cT$, mnemonically indicates a {\em time} centering filter.

\subsection{Frequency Centering, \cF}
\label{sec:freq-centering}
A frequency centering filter subtracts frequency means from a
spectrogram, thus operating in the opposite direction of time centering.  However,
frequency centering is implemented on successive time windows of a
spectrogram to limit the memory requirements for filtering a long
stream of spectra.  Let the $n$-th time window of the $i$-th
spectrogram be
\[
\xim{in} = \left[ \xit{i}{nw-w+1}, \ldots, \xit{i}{nw} \right].
\]
where $w$ is the number of time samples in the window.  The $i$-th
frequency-centered spectrogram is formed by removing frequency means
from successive time windows:
\[ 
\cF \xim{i} \equiv \left[ \xim{i1}\left(\Ident_w - \textstyle\frac{1}{w}\Onemat_{w}\right), \ldots,
                          \xim{iN}\left(\Ident_w - \textstyle\frac{1}{w}\Onemat_{w}\right) \right]
\]
where $N$ is the number of time windows in the spectrogram.  This
operation can be parallelized independently over the frequencies.

\subsection{Energy Clipping, \cC}
\label{sec:energy-clipping-ct}

Signals showing sudden spikes in the energies at all frequencies are
likely to be terrestrial in origin.  
Time centering removes a uniform increase in energy at all
frequencies.  Energy clipping can ameliorate noisy spike that affect
some frequencies more than others by truncating the $L_2$ norm of a spectrum.
This has the advantage of leaving
the large majority of spectra unaltered; only spectra with unusually
large $L_2$ norm are modified.

Energy clipping of a spectrogram is defined as 
\[
 \cC \xim{i} \equiv \left[\phi (\xit{i}{1}), \phi (\xit{i}{2}), \ldots, \phi (\xit{i}{T}) \right]  
\]
where a clipped spectrum is given by 
\begin{equation*}
\phi(\xit{i}{t}) \equiv
  \begin{cases}
    \xit{i}{t}, & \text{if } \vectLength{\xit{i}{t}} < K \\[1ex]
    \displaystyle\frac{K \xit{i}{t}}{\vectLength{\xit{i}{t}}}, & \text{otherwise}.
  \end{cases}
\end{equation*}
with $\vectLength{\xit{i}{t}}=[x_{i1}^2(t)+\cdots+x_{iF}^2(t)]^{1/2}$.


If the elements of $\xim{i}$ are independent standard
Gaussian variates, then $\vectLength{\xit{i}{t}}^2$ is a chi-squared
random variable with $F$ degrees of freedom.  Setting $K^2$ equal to
an upper tail quantile of this distribution allows \cC to pass 
the large majority of input vectors unchanged, truncating 
only a small
fraction of nominally-generated vectors as well as any outlying vectors
whose $L_2$ norm is too large.

\subsection{Huber Normalization, \cH}
\label{sec:huber}
Huber estimation is well-known in robust statistics
\citep{huber1964robust,huber2009robust}.
The idea is to estimate a mean, variance, or
other property of a distribution using thresholded versions of extreme
values so that extreme values cannot unduly influence the estimates.
The Huber calculations produce thresholded residuals on a normalized
scale so residuals have mean approximately zero and standard deviation
approximately one.  These are called {\em winsorized residuals}.  The
recursive Huber filter described below produces winsorized residuals
that are well-suited to detection of dispersed impulses because they
are normalized and they mitigate RFI in individual energy values while
retaining a substantial portion of an impulsive signal.

The Huber filter uses the Huber thresholding function at its core:
\begin{equation*}
  \psi(z) \equiv \left\{
  \begin{array}{rcc}
    -L, & \text{if} &z< -L \\
    z, & \text{if} &-L\leq z\leq L \\
    L, & \text{if} &z> L \\
  \end{array} \right.
\end{equation*}
where the threshold $L>0$ must be specified.  If $z$ is a standard
Gaussian random variable and $L=2$, for example, then $\psi(z)=z$ with
probability about 0.95 and only the 5\% most extreme values of $z$ are
truncated to $\pm L$.

%

For a generic sequence of real values $\yv = [y(1), y(2),
\ldots y(T)]$, the recursive Huber residuals, mean and variance are computed
as 
\begin{align*}
 r(t) &= \psi \left(\frac{y (t )-m( t-1 )}{s (t-1 )} \right), \\[.75ex]
m(t) &= m(t-1) + ps (t-1 )r(t), \\[.75ex]
s^2(t) &= (1-q)s^2(t-1) + (q/c)s^2(t-1)r^2(t),
\end{align*}
for $t=1,2,\ldots,T$.  We initialize the recursion with $m(0) = y(1)$
and $s^2(0) = 1$ and if $s^2(t)$ becomes numerically zero, we
arbitrarily reset it to 1.  Only a long sequence of exactly constant input
data would force a reset.  The value of $c$ is given below along with
some discussion related to choice of tuning constants $p,q\in (0,1)$.

The residual, $r(t)$, is formed by
first standardizing the data value $y(t)$ using mean and
standard deviation estimates from the previous time period and then
applying the $\psi$ function to truncate to $\pm L$, if necessary.  The mean estimate,
$m(t)$, is modified from its previous value by a fraction $p$ of the
rescaled residual, $s(t-1)r(t)$.  In the usual case with
$|r(t)|< L$, the $\psi$ function does nothing and thus
$s(t-1)r(t) = y(t)-m(t-1)$ and the mean update becomes
$m(t) = (1-p)m(t-1)+py(t)$, a weighted average of the previous mean
estimate and the new data value---the usual update formula for an
exponentially weighted moving average (EWMA).  The Huber mean estimate is
thus an EWMA modified by truncation of large residuals.
In a similar fashion, the variance is estimated as a weighted sum
of the previous estimate and the new squared residual.  The constant
$c$ is the expected value of $\psi^2(z)$ with $z$ being a standard
Gaussian random variable, so that $r^2(t)/c$ has
expectation near unity, making $s^2(t)$ nearly unbiased.
Based on the variance of a truncated Gaussian distribution 
\citep[e.g.,][section 10.1]{johnson1994continuous},
we obtain
\[ c = 1 - 2[L\phi(L) - (L^2 - 1)\Phi(-L)], \]
where $\phi$ and $\Phi$ are the standard Gaussian density and cumulative
probability functions.

Choices for $p,q\in (0,1)$ can be made by noting that the EWMA
corresponding to the Huber filter gives weight $p(1-p)^i$ to the
lagged observation $y(t-1-i)$.  This exponentially decreasing sequence
of weights has effective window width $(2-p)/p$ so that an 
effective window width of $n$ samples is obtained by taking
$p=q=2/(n+1)$, and this holds approximately for the Huber analog of the EWMA.

We define winsorized residuals produced by Huber filtering of a sequence
\yv and a spectrogram $\xim{i}$ as
\[
\cH \yv \equiv \left[ r(1), r(2), \ldots, r(T) \right]
\]
and 
\[
 \cH \xim{i}\equiv \left[
\begin{matrix}
\cH \xif{i}{1} \\ \cH \xif{i}{2} \\ \vdots \\ \cH \xif{i}{F}
\end{matrix}
  \right]
\]
respectively.  The Huber filter operates independently over all
frequencies in parallel.

\subsection{Adaptive Interference Cancellation, \cA}
\label{sec:AIC-ca}

\cite{1998AJ....116.2598B}
achieved attenuation of 72 dB 
in the first AIC system used in the radio
astronomy domain.  The value of using more than one reference antenna
was also made explicit: it permits cancellation of a greater number
of uncorrelated noise sources.  We use the primary idea of
AIC---subtracting interference identified by correlating with
reference antennas---but adapt it to the fast spectrum data collected
by ATA in Fly's Eye mode.  The major difference from typical
applications is that phase information is not available in the ATA
energy spectra.  Even so, interference can be estimated by dynamic
linear combinations of the primary and reference spectra.  Subtracting
estimated interference from the primary spectra mitigates its impact
on dispersed pulse detection.

AIC assumes that a primary sensor receives a signal of interest plus
interference that is not correlated with the signal.  In addition,
some number of reference sensors do not receive the signal of interest
but do receive the same interference as the primary sensor plus noise
that is uncorrelated with the interference.  These are reasonable
assumptions for Fly's Eye data because individual antennas are sensitive to
astronomical signals originating in different regions of the sky and both atmospheric
and terrestrial RFI typically infect multiple antennas.

An important quality of AIC is that it makes very few assumptions
about the characteristics of the interference.  AIC can be used to
mitigate a wide class of RFI that appears simultaneously in multiple
antennas.  If a specific RFI pattern is prevalent (e.g., the 60 Hz
signals common in radio astronomy observations), then an effective RFI filter can be designed to mitigate
that specific type of signal.  AIC is likely to be somewhat less
efficient at mitigating specific known interference patterns, but AIC
is adaptive in the sense that it will handle a wide class of
interferers without the need to design a different filter for each
type of interference.

The AIC filter on a spectrogram is defined on successive windows of an
input sequence $\xif{i}{f}$.  Let the $n$-th window of $\xif{i}{f}$ be 
\[
\xv_{ifn}=\left[ \xift{i}{f}{(n-1)w+1}, \ldots, \xift{i}{f}{nw} \right].
 \]
AIC is first applied to $\xv_{if1}$, then to $\xv_{if2}$, and so on.

To filter the $n$-th window, create a matrix, $\Am_{ifn}$,
with rows being the windowed signals for frequency $f$ from each 
of the reference spectrograms, $\xv_{kfn}$, $k \in \sK_i \subseteq
\{1,\ldots,i-1,i+1,\ldots,I\}$.  For example, 
if the reference set for cleaning $\xv_{1fn}$ is
$\sK_1=\{2,\ldots,I\}$ (i.e., all other signals), then 
\[
\Am_{1fn} = \left[
\begin{matrix}
\xv_{2fn} \\
\xv_{3fn} \\
\vdots \\
\xv_{Ifn} \\
\ov_w'
\end{matrix}
\right]
\]
where $\ov_w$ is a vector of $w$ ones and is included in every $\Am_{ifn}$ matrix.

AIC cleans the signal $\xv_{ifn}$ by subtracting its
linear projection on $\Am_{ifn}$ to obtain residuals
\[
\ev_{ifn} \equiv \xv_{ifn}(\Ident_w - \Am_{ifn}'(\Am_{ifn}\Am_{ifn}')^{-1}\Am_{ifn})
\]
where $\Ident_w$ is the identity matrix of dimension $w$.  For
a signal $\xv_{if}$ consisting of $N+1$ windows, AIC cleaning is
denoted by
\[ \cA \xv_{if} = [\ev_{if0}, \ldots, \ev_{ifN}] \]
and parallel cleaning of each frequency in a fast spectrogram is denoted by 
\[
 \cA \xim{i}\equiv \left[
\begin{matrix}
\cA \xif{i}{1} \\ \cA \xif{i}{2} \\ \vdots \\ \cA \xif{i}{F}
\end{matrix}
  \right].
\]

In an antenna array, different antennas will have different frequency
response curves and will receive different RFI signals due to
differences in direction of arrival, local terrain, and so forth.  The
AIC formulation is effective, however, even when the intensity of RFI
varies by antenna and by frequency because the projection onto
$\Am_{ifn}$ is the best (least-squares) linear combination of the
reference windows for estimating the RFI in the target input sequence $\xv_{ifn}$.

The algorithm requires specification of the window width $w$ and the
set of reference signals.  We set $w=640$ samples, which, at the
sampling rate of 1600 Hz, is about 400 ms.  Choice of reference
signals is discussed in Section~\ref{sec:methods}.

Note that for the Fly's Eye data we are operating on power spectra
rather than time-series voltage data, which is the conventional target.  
Nevertheless, the projection method described
here applies to both and can be used to remove RFI.

\section{Chirp Detection }
\label{sec:detect}
\begin{figure}[t]
 \begin{center}
 \includegraphics[width=0.6\textwidth]{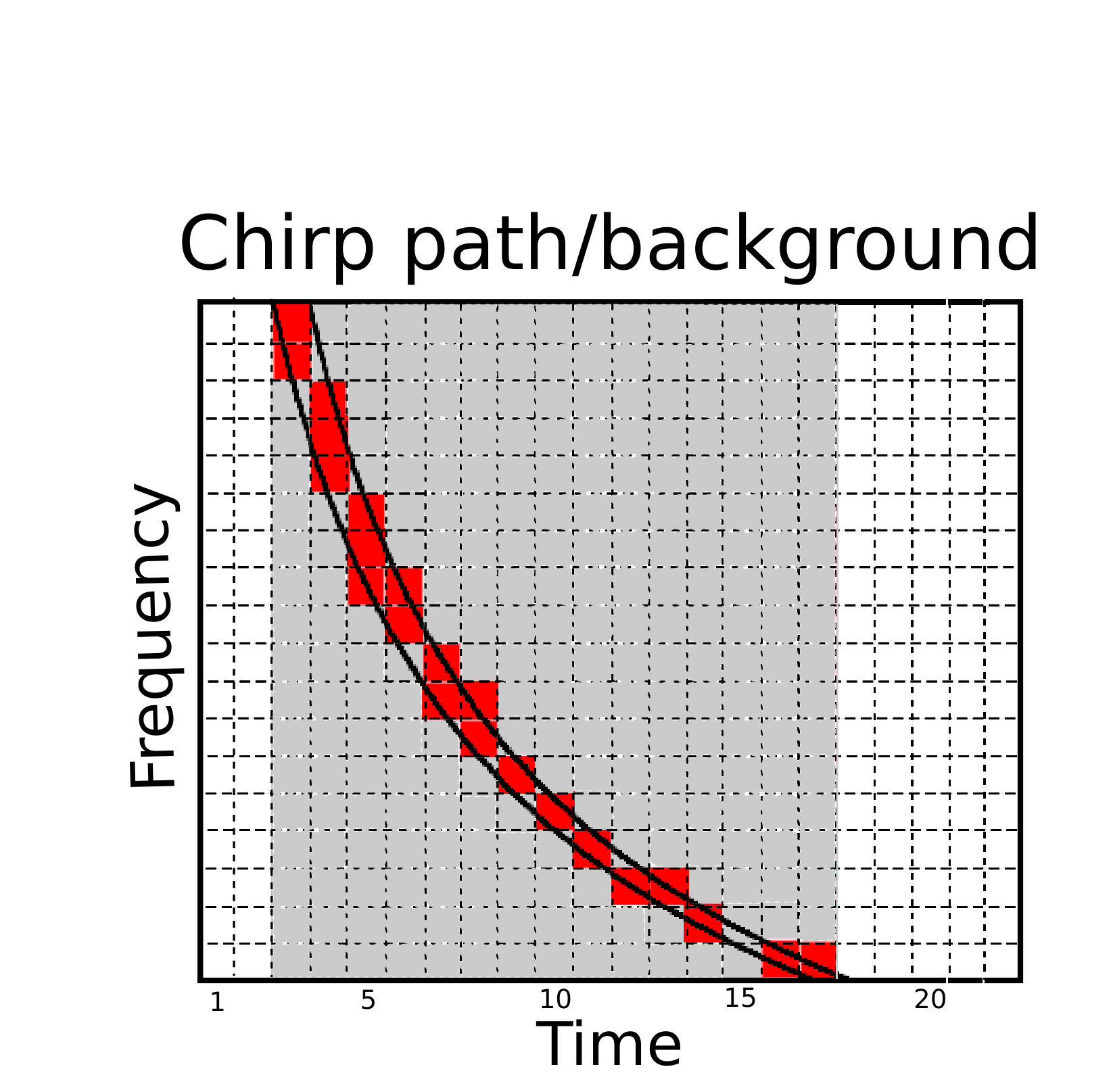}
\end{center}
\caption{The chirp path for a chirp starting at time 3 is shown by the red pixels.  The two bold curves illustrate chirps with the same dispersion measure.  One chirp starts at the beginning of time sample 3 and the second chirp starts at the end of time sample 3.  The pixels in the chirp path are those pixels with centers between the two continuous chirps.}
\label{fig:chirpPath} 
\end{figure}

Examples of chirped signals from the Crab pulsar appear
in Figure~\ref{fig:crabs}.  The presence of a chirp is indicated by
higher than expected values of energy in the pixels along the
\textit{chirp path}.  We define the chirp path with time index $t$ and
dispersion measure \DM as
\begin{align*}
  \sC(t,\DM)= \{ (f,\tilde{t}): 
  & \text{ the center of pixel $(f,\tilde{t})$ is bracketed by chirps that} \\
    & \text{begin at $t\pm 0.5$ with dispersion measure \DM}  \}.
\end{align*}
This is illustrated in Figure~\ref{fig:chirpPath} for $t=3$ and a
specific dispersion measure.  The black curves are chirps starting at
$t\pm0.5$ with dispersion delay given by the cold plasma dispersion law.
The grid of dashed lines indicates pixel boundaries in a
spectrogram.  If the center of a pixel lies between the chirps then it
is in the chirp path for the given start time ($t$) and dispersion measure (DM).
%
%

While there are many other ways that a discrete chirp path could be defined, this
definition has the desirable quality that every pixel in the
spectrogram belongs to one and only one chirp path for a given
dispersion measure.  Therefore,
successive chirp paths do not share pixels and so are statistically independent of
each other as long as the energy values at different pixels are independent of each
other.

\subsection{Sequence of T-tests}
\label{sec:t-test}
Our approach to determining whether a chirp is present along a chirp path is to perform a
standard t-test comparing the mean energy for pixels in the chirp path
to the mean for pixels in the background.  The set of background pixels for a chirp
starting at time $t$ with dispersion measure $\DM$ is denoted as
$\bar\sC(t,\DM)$ and consists of pixels concurrent with the chirp
path but excluding the chirp path, as indicated by gray shading in Figure~\ref{fig:chirpPath}.

The t-test and the
effects of deviations from its sampling assumptions are discussed in many
introductory statistics tests \citep[e.g.,][]{glass1996statistical}.
The Gaussian assumption becomes less important with larger samples.
Non-constant means and variances in the chirp path or the background
have the potential to affect both the false alarm rate of the 
test statistic and its ability to make correct detections.

A large t-score indicates the presence of a chirp.  If the pixels in
$\sC(t,\DM)$ and $\bar\sC(t,\DM)$ are independent and identically
distributed (iid) samples drawn from Gaussian distributions,
then the t-scores are drawn from a t-distribution with
$n_b+n_c-2$ degrees of freedom. 

In this paper we only explore analysis of pulses with a width equal to the sampling time.  
Real astronomical pulses, however, can have widths much larger than the sampling time.
These broader pulses can be addressed through hierarchical averaging of the time series data
and application of the same chirp integration.

\section{Tests on Synthetic Chirps}
\label{sec:methods}
We tested various combinations of the RFI mitigation filters described in
Section \ref{sec:RFI} on ATA data in Fly's Eye mode.  Specifically,
one hour of power spectrograms was extracted simultaneously from 44
receivers with 128 frequency channels from $\nu_1 = 1325$ MHz to
$\nu_F = 1535$ MHz, with a frequency resolution of $\Delta f =1.64$
MHz per channel.  (Although the ATA has 42 antennas, each has a dual linear polarization feed, making 84 independent signal streams. Our hardware captures 44 of these
signals, and we refer to them as antennas to simplify exposition.)

\begin{figure}[thp]
 \begin{center}
   \includegraphics[width=0.495\textwidth]{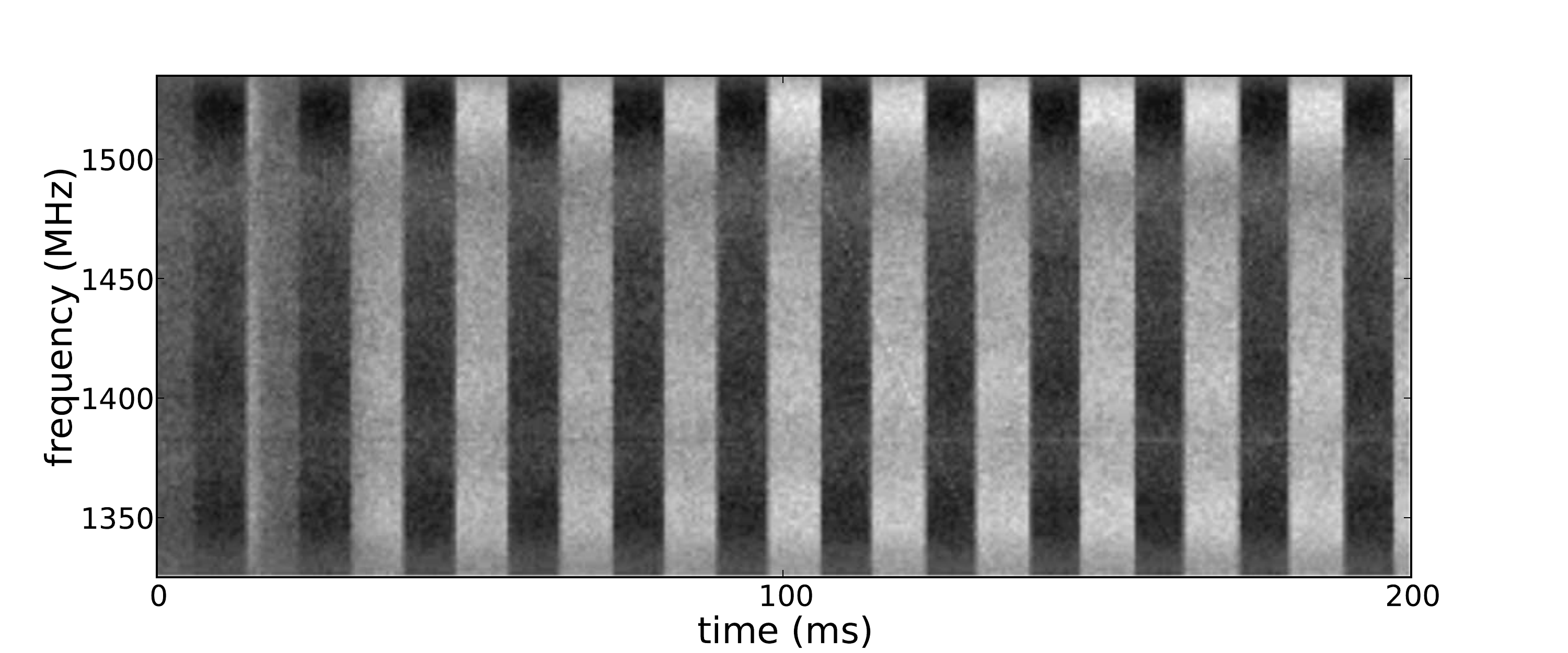}
   \includegraphics[width=0.495\textwidth]{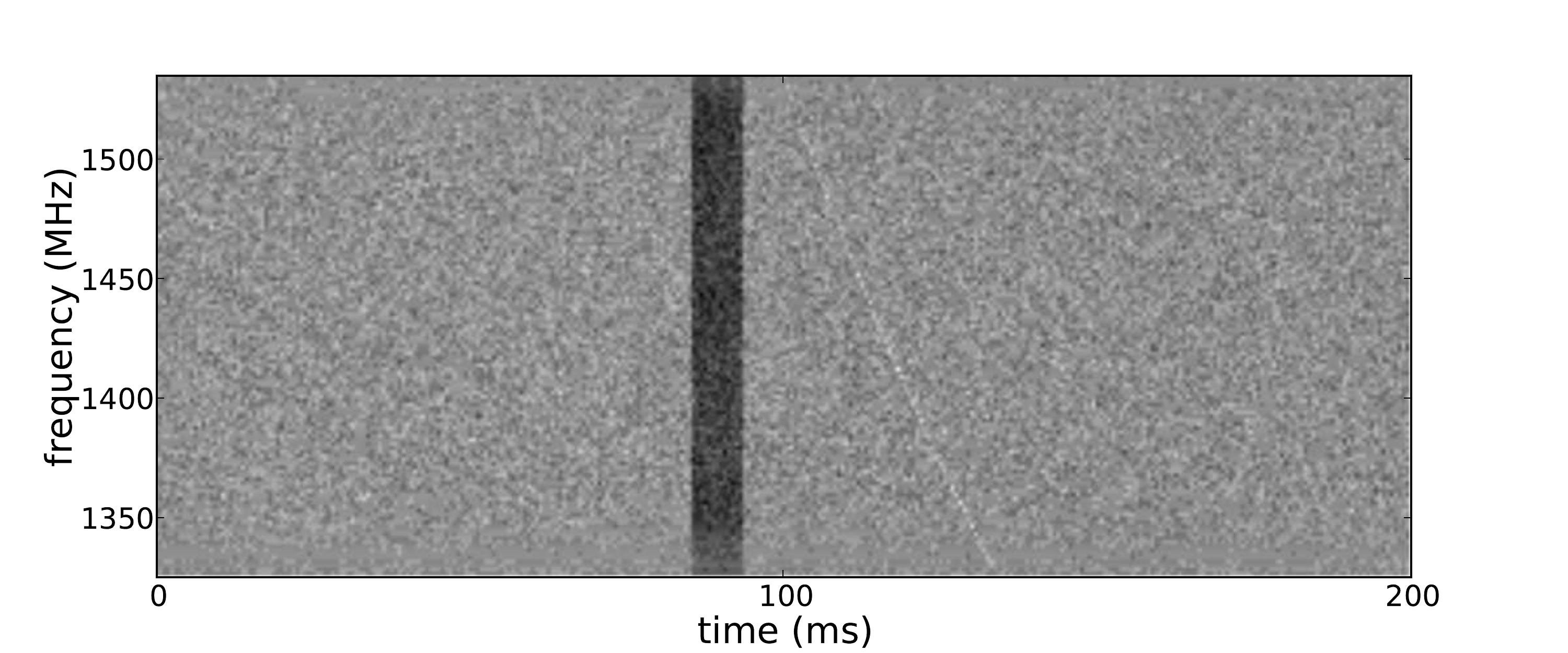} \\[-1.5em]
   (A) \hspace{.44\textwidth} (F) \hfill \ \ \\
   \includegraphics[width=0.495\textwidth]{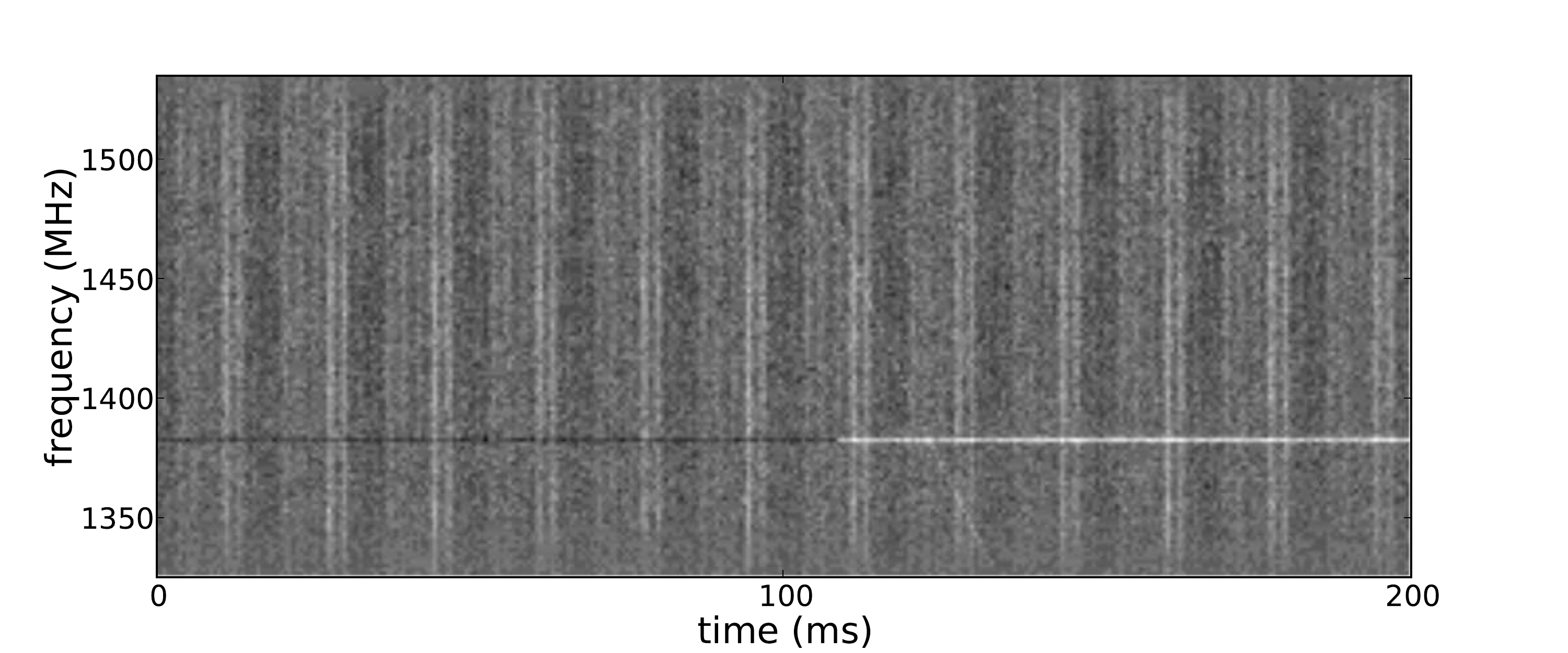}
   \includegraphics[width=0.495\textwidth]{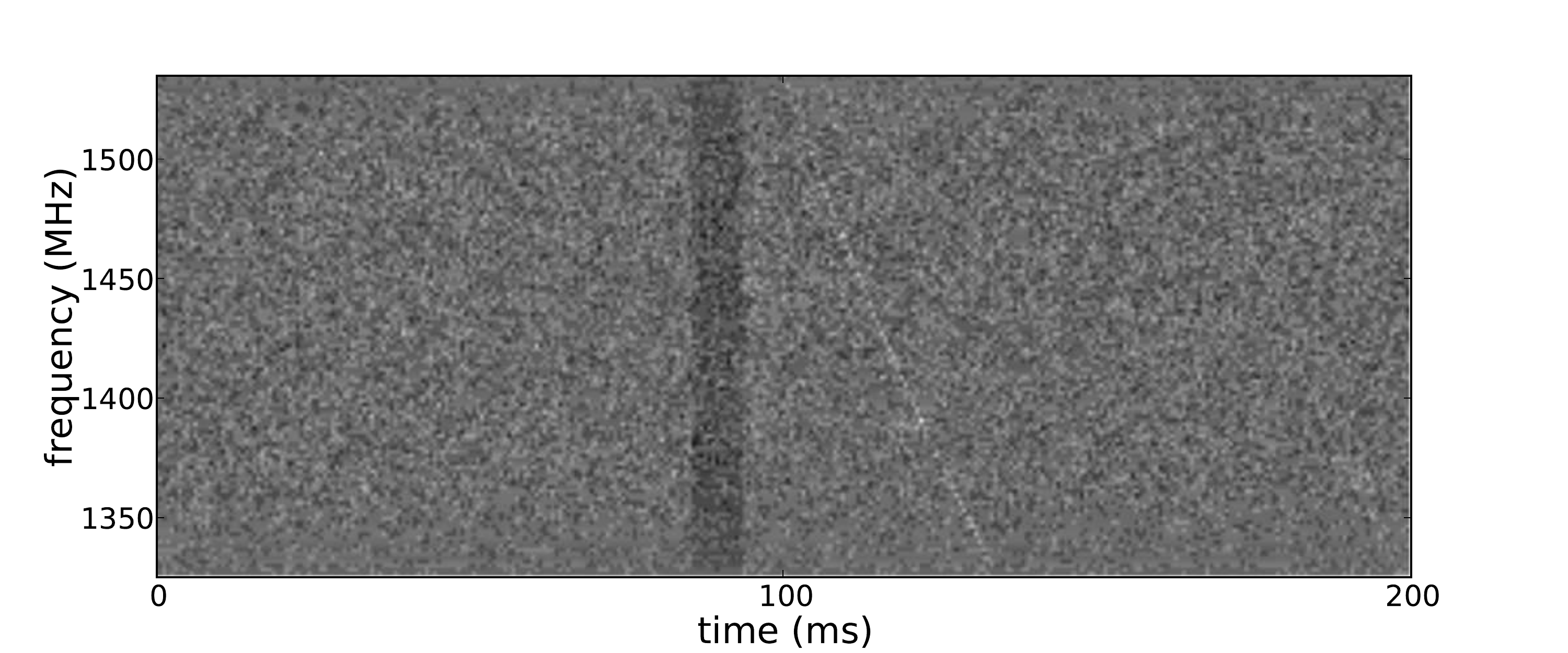} \\[-1.5em]
   (B) \hspace{.44\textwidth} (G) \hfill \ \ \\
   \includegraphics[width=0.495\textwidth]{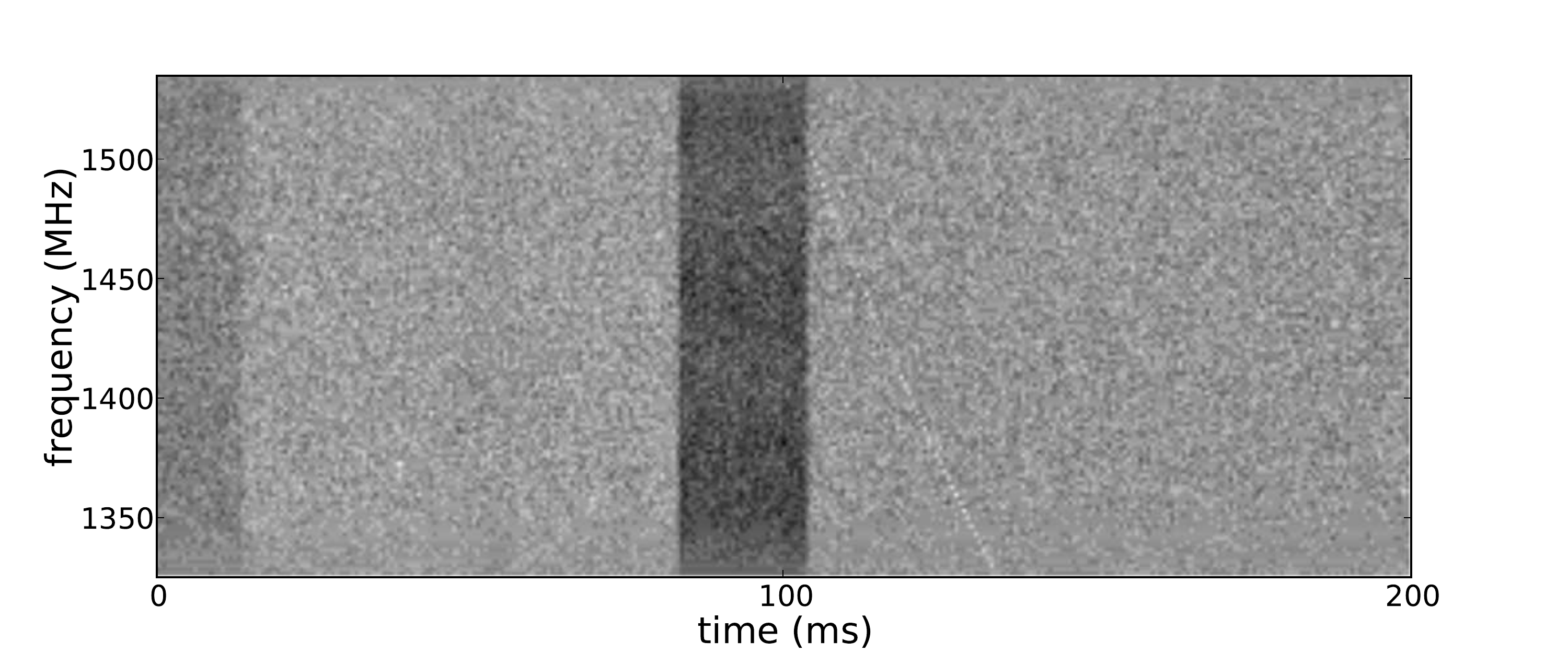}
   \includegraphics[width=0.495\textwidth]{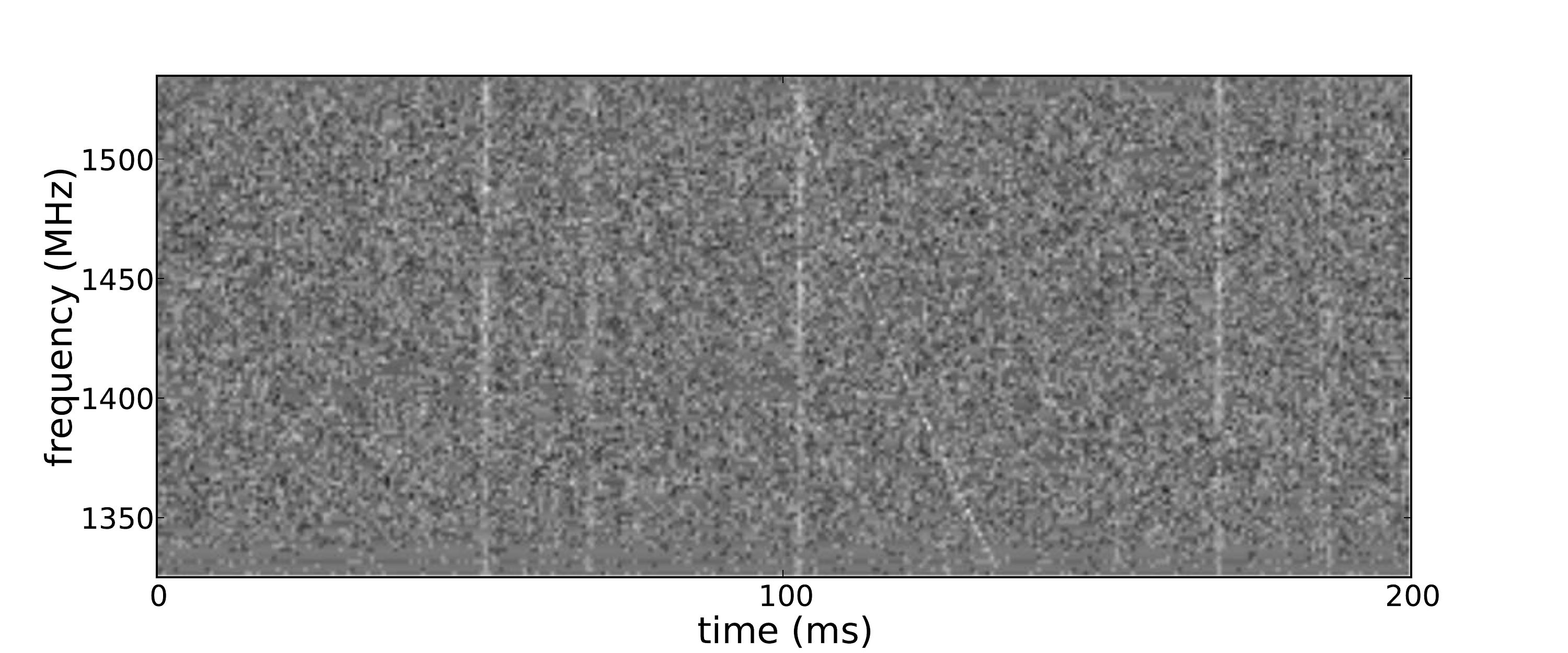} \\[-1.5em]
   (C) \hspace{.44\textwidth} (H) \hfill \ \ \\
   \includegraphics[width=0.495\textwidth]{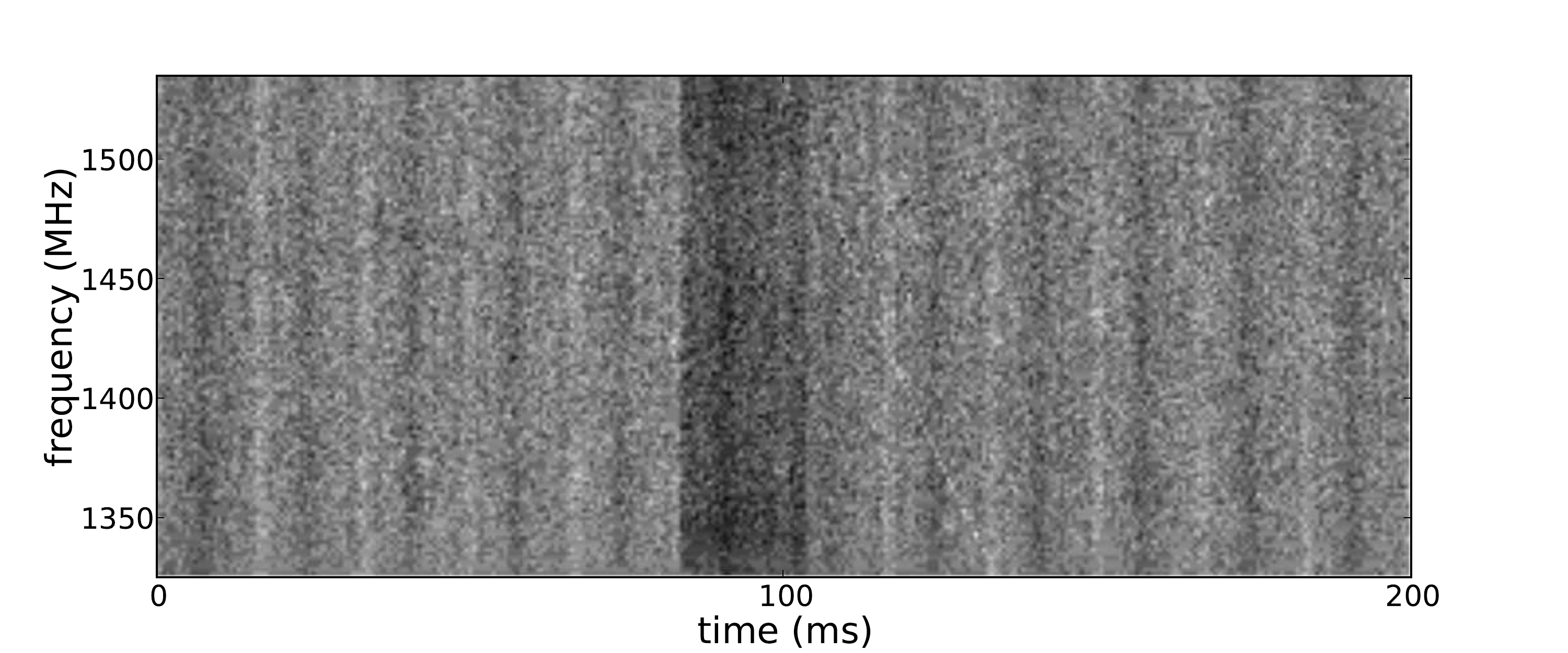}
   \includegraphics[width=0.495\textwidth]{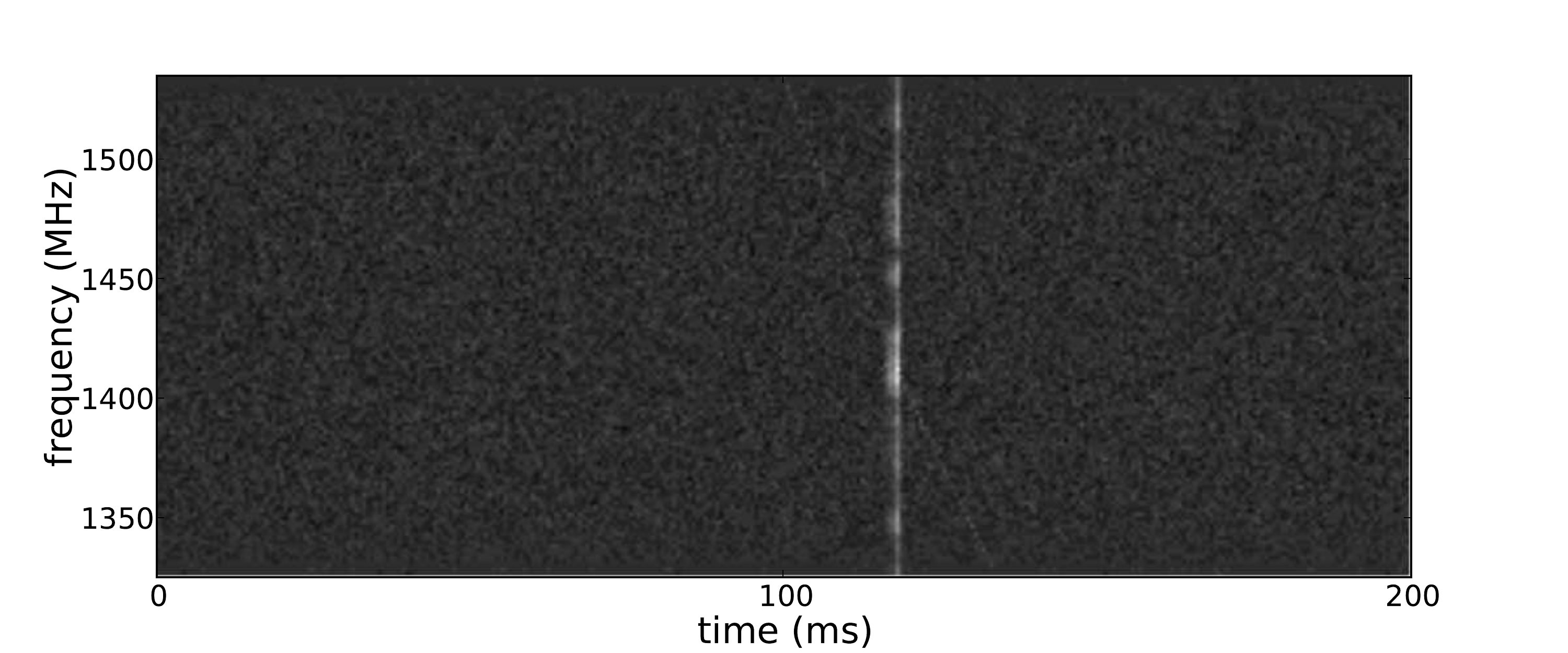} \\[-1.5em]
   (D) \hspace{.44\textwidth} (I) \hfill \ \ \\
   \includegraphics[width=0.495\textwidth]{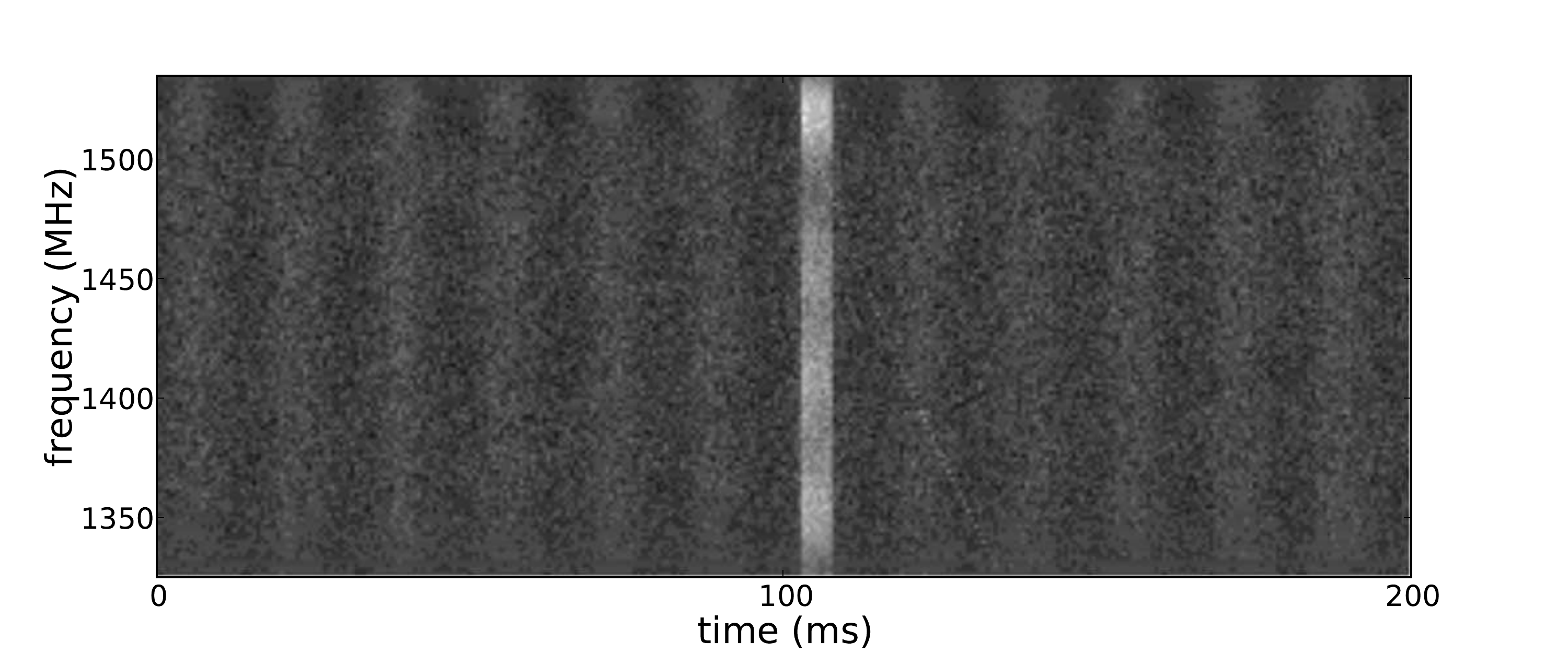}
   \includegraphics[width=0.495\textwidth]{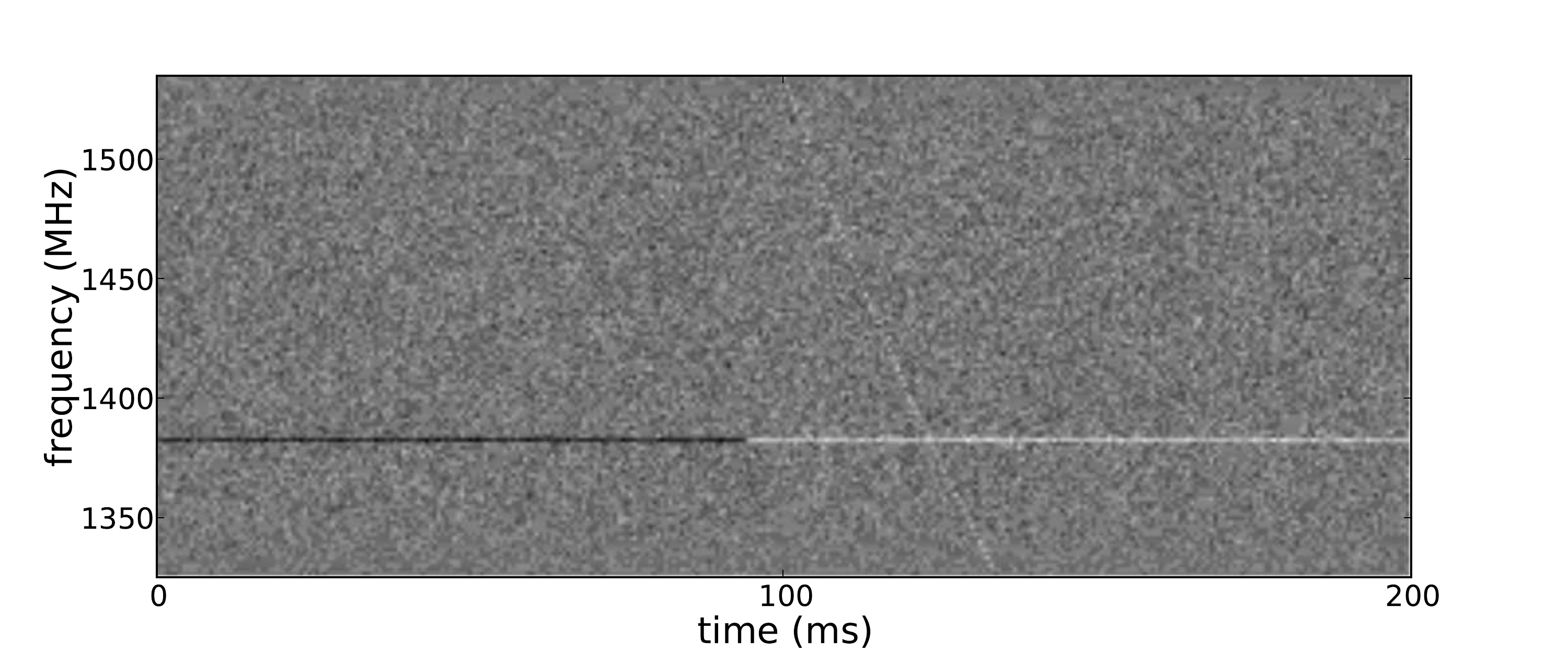} \\[-1.5em]
   (E) \hspace{.44\textwidth} (J) \hfill \ \ \\
\end{center}
\caption{Ten examples of RFI, (A)--(J), with chirps embedded beginning at 100 ms at an energy level of 2.0.}
\label{fig:embed} 
\end{figure} 

Figure~\ref{fig:embed} shows a ``Rogues Gallery'' of 10 spectrogram
segments containing different amounts and types of unwanted signal.
These are some of the worst cases.  The vast majority of data segments
would appear as white noise in this type of display.
Each spectrogram in Figure~\ref{fig:embed} also has an
artificially embedded chirp that is visible beginning at
100 ms in the highest frequency of each segment and sweeping through
the frequency band over about 35 ms of time.  These 10 segments
include impulsive RFI, such as the white vertical lines in (H) and
(I), as well as single-frequency transients, such as the horizontal
streaks at about 1375 MHz in (B) and (J).  The black and white bands
in (A) alternate at 60 cycles per second, which is likely an instrumentation
artifact, along with the similar but subtler patterns in (B), (D) and
(E).  We refer to all of this clutter as RFI, even though some of it
is not likely to have originated in the radio frequency domain.

The artificial chirps in Figure~\ref{fig:embed} have a $\DM=57\text{pc/cm}^3$ corresponding to that of the Crab pulsar.  To
embed a chirp starting at a particular time, we added energy to
appropriate frequency bins in each spectrogram during the duration of
the chirp.  We use the cold plasma dispersion law to find the chirp frequency
for each time sample from the start of the chirp to the end of the
chirp.  For each time sample during a chirp, energy was typically
added to two frequency bins -- the two bins bordering the chirp
frequency---with the total energy $E_0$ split in proportion to the distances between the bin frequencies and the chirp frequency.  For example, if the embedded chirp has frequency $\nu_c$ and the bordering bins have center frequencies $\nu_L$ and $\nu_H=\nu_L+Delta$, then energy $E_0-E_0(\nu_c-\nu_L)/\Delta$ is added to the bin with frequency $\nu_L$ and the remainder is added to the next higher bin.


The examples in Figure~\ref{fig:embed} show embedded chirps with
energy level $E_0 = 2.0$.  This is the strongest of four energy levels
used for this study: 0.75, 1.0, 1.25 and 2.0.  (Although measurements
in our spectrograms are proportional to energy, at the time these data
were collected the ATA Fly's Eye mode did not calibrate each antenna
to an absolute reference.  Therefore, embedding chirps with fixed
energy $E_0$ into uncalibrated data equates to embedding signals of
differing absolute strength, depending on the antenna gain.)
\begin{figure}[p]
 \begin{center}
   \includegraphics[width=\textwidth]{completeibob09Dchrp10.pdf} \\
   \includegraphics[width=\textwidth]{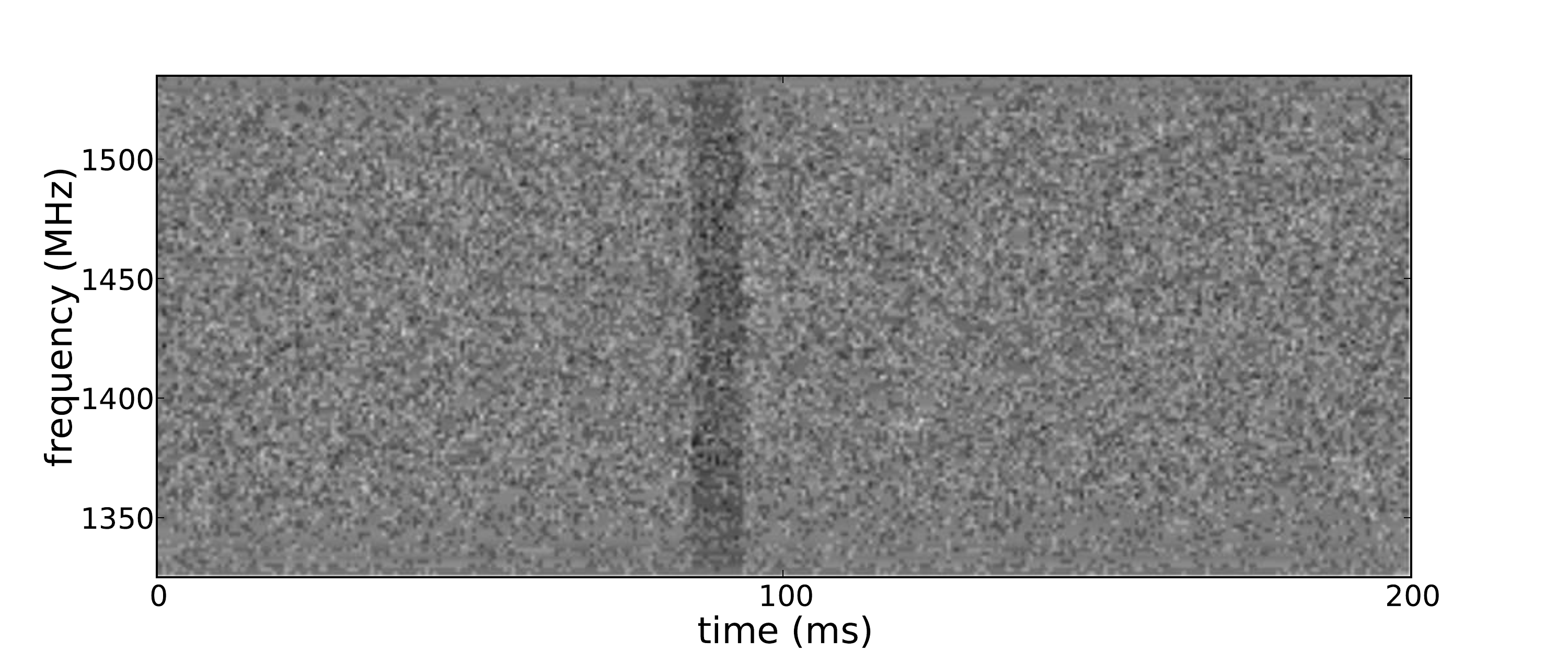} \\
\end{center}
\caption{Strongest ($E_0=2.0$) and weakest ($E_0=0.75$) embedded energy on the (J) section of RFI.}
\label{fig:embed2} 
\end{figure} 
Figure~\ref{fig:embed2} shows the RFI in panel (G) from
Figure~\ref{fig:embed} with the chirp embedded at the strongest
($E_0=2.0$, top) and weakest ($E_0=0.75$, bottom) energy levels.  The
weak chirp is barely visible in the lower image but was detected at a
level as low as 5 false alarms in 138 million time samples, equivalent to one full day of observing.

\begin{figure}[t]
 \begin{center}
   \includegraphics[width=\textwidth]{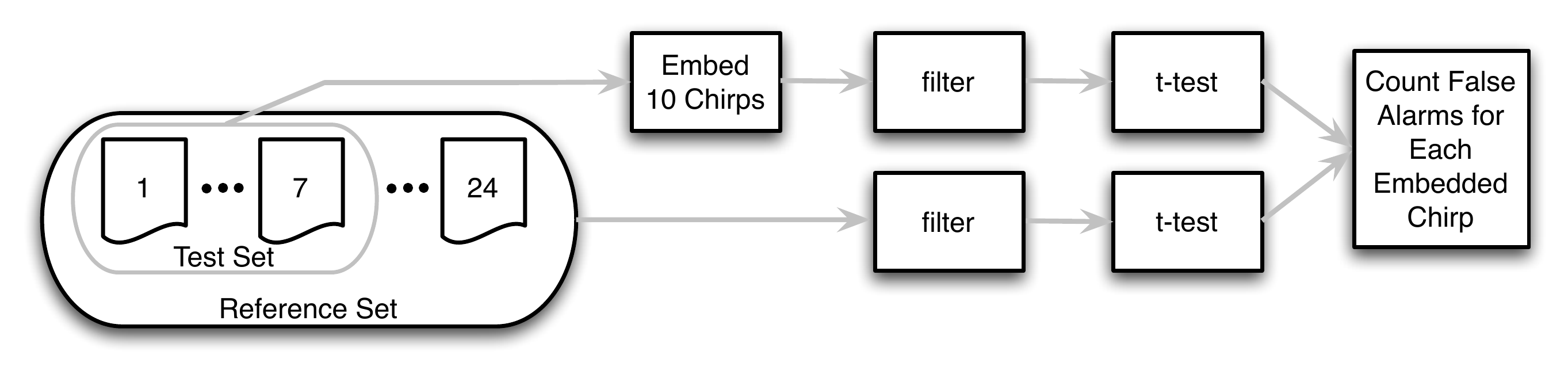} \\
\end{center}
\caption{Process of testing a detection method on synthetic chirps.  The figure illustrates the processing that was done for a given embedded signal strength, $E_0$, and a given RFI filter. This was repeated for each combination of four signal strengths and six filters.}
\label{fig:flow} 
\end{figure}

Figure~\ref{fig:flow} illustrates the process we used to compare the
effectiveness with which various filtering strategies (discussed further below) mitigate RFI.
Each of the 44 antenna output files used for our study is
approximately an hour in length.  Seven of these files were used to
provide examples of RFI. In three of the seven files, we
studied RFI at two different times, and in the remaining
four, we chose only a single time, making total of 10 different 
RFI exemplars as shown in Figure~\ref{fig:embed}.

Chirps were embedded with each of the four energy levels listed
above in each of the 10 exemplars for a total of 40 examples
of embedded chirps.  The data set with embedded chirps is referred to as
the {\em test set}.

Another 17 of the 44 files were randomly selected, so that together
with the seven files in the test set, a total of 24 files are identified
as the {\em reference set}.  The reference set represents approximately 24 hours worth
of data. The reference set was processed identically to the test set
and used to determine how many false alarms to expect per day for
thresholds corresponding to detection of each of the 40 different
embedded signals.

From the remaining 20 files (those not used in the reference set), 10
were randomly chosen
as the cleaning signals to use with AIC filtering.
The remaining 10 files were not used in the tests reported here.

Each filtering strategy aimed at mitigating RFI was applied to both
the test set and the reference set, followed by t-test calculations
for chirp detection.  The number of t-scores in the reference set
greater than or equal to the t-score for each embedded chirp in the
test set is the number of false alarms.  False alarm counts were
computed for each combination of four energy levels and the 10 RFI
sections in the reference set, giving 40 false alarms counts for a
given filtering strategy.

Six filtering strategies were tested using the processing described
above and illustrated in Figure~\ref{fig:flow}.  Each filtering
strategy consists of a combination of the mitigation filters presented
in Section~\ref{sec:RFI}.  The six strategies, identified by short
names, are defined below and rationale is given for the specific
choices selected for each strategy.  Frequency centering and AIC
filters operate on consecutive time windows of spectrograms.  In each
case below, the window size was set to $w=640$ samples, corresponding
to 400 ms time windows.  This window size is about 11 times longer
than the embedded chirps so that the chirp represents only a small
fraction of data in a window.

The filtering strategies are as follows:
\begin{description}
\item[none:] is the t-test detector on the raw spectrograms---no filter is applied.  This is included for reference, even though it is not a competitive strategy. 
\item[center freq:] is frequency centering in which frequency means
  are subtracted from successive spectrogram windows.  This filter is
  included primarily for a comparison with the following strategy.
\item[center freq + AIC:] applies frequency centering to the file to
  be cleaned and to each of the 10 AIC cleaning files.  Then AIC is
  run on each window of the spectrogram.  Comparing this strategy to
  the previous one allows for measuring the value of the more complex
  AIC calculations beyond the simple frequency centering calculations.
If no reference signals were used in the
AIC cleaning algorithm, AIC would be identical to the "center freq" approach above.
To emphasize
that AIC centers the frequencies \emph{and} uses additional reference signals, we
write "center freq + AIC" for the case where frequency centering is applied to the file
to be cleaned and to each of the 10 AIC cleaning files before applying AIC.
\item[center freq,time:] performs frequency centering (in windows) and
  then time centering.  This allows for comparing the value of time
  centering beyond that of frequency centering (the second strategy)
  and provides a reference for the following strategy.
\item[center freq,time + AIC:] is frequency and time centering (as
  above) on the file to be cleaned and on each file in the AIC
  cleaning set.  Then AIC is run on each window of the spectrogram.
  Comparison to the previous strategy shows the incremental value of
  AIC beyond two-way centering.
\item[Huber + center time + energy truncation:] first applies Huber
  filtering with $p=q=0.001249$, and $L=2$.  Next the mean energy at
  each time is removed. Finally the energy clipping filter is applied
  with a threshold of $K=155$.  This strategy is intended to determine
  the value of outlier mitigation as implemented in Huber filtering
  and energy clipping.  Choices of $p$ and $q$ come from setting the
  effective window width to 1600 samples ($p=q=2/1601$), so that the
  smoothing is scaled to one second of time.  This is much longer than
  the 400 ms used for AIC and frequency centering.  The rationale is
  that the combination of Huber filtering and energy clipping
  mitigates outlying RFI data values so that smoothing can be done on
  longer time scales without the risk of bad data corrupting the
  filter output for a prolonged period.  The value $L=2$ should clip
  approximately 5\% of the individual data values in the Huber
  thresholding function.  Similarly $K=155$ is the 0.95 quantile of
  the $\chi^2$ distribution with $F=128$ degrees of freedom and this
  should clip energy from approximately 5\% of the normalized spectra
  as explained in Section~\ref{sec:energy-clipping-ct}.  These are
  reasonable choices informed by usual practices in application of
  robust statistics.
\end{description}

\section{Results \& Discussion}
\label{sec:results}

\subsection{Detection Sensitivity and False Alarm Rates}
\label{sec:detect-sens}

A good measure of the ability of a filter combination to mitigate RFI
and retain the signal of a chirp is the count of t-scores greater than
or equal to the t-score of a given chirp.  For example, after
embedding a chirp in one of the test-set files and running a given
filter combination over the data, suppose the chirp generates a
t-score of $t_0$.  The number of false alarms for this case is
determined by counting how many t-scores are greater than $t_0$ when
the same filtering is applied to the 24 hour reference set with no
embedded chirps.  False detection counts are obtained for each
embedded chirp in combination with each choice of RFI filter.

\begin{figure}[t]
 \begin{center}
   \includegraphics[width=\textwidth]{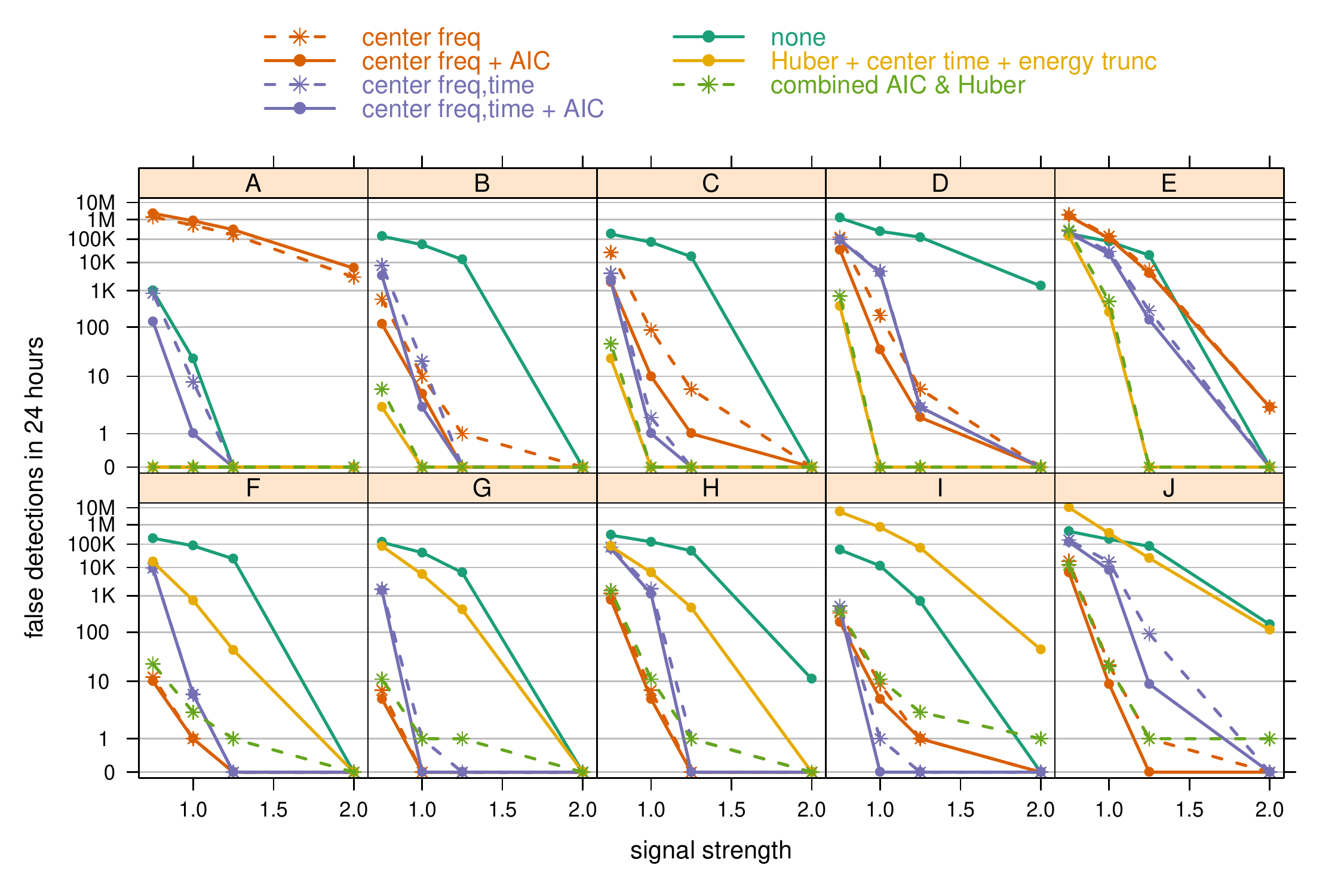}
\end{center}
\caption{False alarm counts on 10 embedded chirps using six
  combinations of RFI mitigation filters.}
\label{fig:alarms} 
\end{figure}

Figure~\ref{fig:alarms} plots false detections in 24 hours
corresponding to each of the 10 RFI examples shown in
Figure~\ref{fig:embed} with various strengths of embedded chirps and
various combinations of filtering.  Within each panel, different lines
correspond to different filter combinations and each line presents
false detections for each of five embedded chirp strengths.  Six of
the seven lines are for the six filter combinations detailed in
Section~\ref{sec:methods}.  The seventh, labeled ``AIC \& Huber,'' is
discussed below.  False detection counts drop toward zero as the
strength of the embedded signal increases.  The vertical axis is
scaled to emphasize the low range of false detections.  Values
above 100 represent chirps that could not reasonably be detected
by close individual analysis of all detections in a single day of observation time.

The RFI examples, A--J, in Figure~\ref{fig:embed} and the corresponding
panels in Figure~\ref{fig:alarms} are ordered
according to decreasing success of the Huber filter, center time,
and energy clipping combination.
Interestingly, centering filters with and without AIC tend to perform
well when Huber does poorly, and vice versa.  In particular, in the
top row of plots, the Huber filter is uniformly best, while it is not
much better than no filtering in the bottom row of plots.  However,
the ``center freq. + AIC'' filter is best in the bottom row of plots,
except for case ``I'' where the two frequency and time centering filters do
somewhat better.

The fact that the Huber filter complements ``center
freq. + AIC'' suggests that a better strategy would be to combine both methods, as described below.
In fact, the lines labeled ``combined AIC \& Huber'' show that the
combined AIC and Huber method does indeed perform well on all ten
embedded chirps.  The combined method is not uniformly best, but it is
the only strategy that never performs poorly relative to the other
filters.

False alarms for the ``combined AIC \& Huber'' strategy were computed
as follows.  Let $C$ denote the minimum of the false detection counts
for ``center freq. + AIC'' and for ``Huber + center time + energy
truncation''.  This $C$-score is computable from the empirical distribution
of $t$-scores over the 24 hour reference set and is itself a detection
statistic.  A small $C$-score near zero denotes strong evidence of a chirp.
For an embedded chirp with $C=c_a$, the false detection count is the
number of $C$-scores in the reference set that are less than or equal to $c_a$.
These counts are shown in Figure~\ref{fig:alarms} for the ``combined
AIC \& Huber'' method.

\section{Summary and Conclusions}
\label{sec:summary-conclusions}

Radio frequency interference is a dominant limiting factor in the design and performance
of fast radio transient experiments.
We present here an analysis of the effectiveness of several RFI mitigation methods, 
with the goal of a more rigorous statistical understanding of the
performance of these filters.  We apply these methods to actual interference present in data obtained 
as part of the ATA Fly's Eye survey.  A search for synthetic dispersed pulses that
 were added to the interference data 
was employed 
as a means to determine the rate of false detections.  Filters explored in various combinations
include time centering, frequency centering,
adaptive interference cancellation (AIC), Huber filtering, and energy clipping.

Huber filtering in combination with energy clipping and time centering proved very effective at eliminating RFI that 
was primarily broad in frequency but variable in time.  In many cases, application of these filters 
led to zero false positives when applied to 24 hours of data.  For the case of RFI that is predominantly frequency-dependent,
the Huber filter is largely ineffective.  In these cases, frequency centering with and without AIC is the most effective.  Unfortunately,
the frequency centering approach produces many false positives for broadband RFI.  A method that combines Huber and AIC
filtering proves uniformly effective over a range of time- and frequency-dependent interferers.  It is the only method explored
that will produce a low number of false positives under all RFI examples considered here.

AIC is computationally intensive and is applicable to only the 
subset of experiments in which reference
signals are available.
The computational complexity of AIC is much higher than that for using a robust filtering approach.  
For example, given signals from $N$ antennas, each having $F$
frequency channels output at each time, we will need to use AIC on
$NF$ primary signals with up to $N-1$ reference signals for each
primary signal.  This implies that we need to form and solve $NF$
systems of linear equations for each primary signal in each window of
time.  If we use the maximum number of reference signals, then the
algorithm does not scale well with increasing numbers of radio
antennas.  While modern computing devices are often optimized to do
linear algebraic calculations as needed for AIC, as the number of antennas grows, the computational burden may eventually be too high.

To decrease the computational costs, compromises will need to be made.  For example, we may have to arrange radio telescopes so that telescopes that are close to each other point to different locations in space.  This will help ensure that the same noise signals are observed by each telescope but the signals of interest will not be observed by multiple telescopes.  Doing this will allow fewer reference signals to be used for each primary signal.  Alternately, AIC could be used as a second stage of processing to further clean spectra that are suspected of having dispersed impulses.  We need to understand the trade-offs of various AIC algorithms, 
and how the AIC algorithms interact with other RFI mitigation techniques as well
as pulse detection techniques.

Finally, we note that the forms of RFI used for testing these algorithms are by no
means exhaustive.  It is certainly of interest to apply these algorithms to data obtained from
other telescopes in different RFI environments.  Nevertheless, we see promise 
in Huber filtering and AIC for mitigation of RFI from incoherent spectra in the search for fast
radio transients.

\acknowledgements

The authors would like to acknowledge the generous support of the Paul
 G. Allen Family
 Foundation, who have provided major support for design, construction,
 and operations of
 the ATA. Contributions from Nathan Myhrvold, Xilinx Corporation, Sun
 Microsystems,
 and other private donors have been instrumental in supporting the ATA.
 The ATA has been
 supported by contributions from the US Naval Observatory in addition
 to National Science
 Foundation grants AST-050690 and AST-0838268.
This work received funding from Los Alamos National Laboratory LDRD
Project \#20080729DR.
We thank Griffin Foster and Peter McMahon for their contributions to the
Fly's Eye experiment.


\end{document}